\documentclass[pre,superscriptaddress,twocolumn,floatfix]{revtex4}

\usepackage{graphicx}

\begin{document}

\title{Clusters in a magnetic toy model for binary granular piles}

\author{K. Trojan} \affiliation{SUPRAS and GRASP\footnote{GRASP = Group for
Research in Applied Statistical Physics;\\ SUPRAS = Services
Universitaires Pour la Recherche et les Applications en
Supraconductivit\'e,  a member of SUPRATECS Centre}, Institute of
Physics, B5, University of Li$\grave e$ge, B-4000 Li$\grave e$ge,
Euroland} \affiliation{Institute of Theoretical Physics, University of
Wroc\l{}aw,  pl. M. Borna 9, 50-204 Wroc\l{}aw, Poland}
\author{M. Ausloos} \affiliation{SUPRAS and GRASP\footnote{GRASP = Group for
Research in Applied Statistical Physics;\\ SUPRAS = Services
Universitaires Pour la Recherche et les Applications en
Supraconductivit\'e,  a member of SUPRATECS Centre}, Institute of
Physics, B5, University of Li$\grave e$ge, B-4000 Li$\grave e$ge,
Euroland}

\begin{abstract}
Results on a generalized  magnetically  controlled ballistic deposition
(MBD) model of granular piles are reported in order to search for the
effect of  ''spin flip''   probability $q$  in building a granular pile.
Two different regimes of spin cluster site distributions   have been
identified, a border line $q_c(\beta J)$ where $J$ is the interaction
potential strength.
\end{abstract}

\maketitle

\section{ INTRODUCTION}

The physics of granular matter has drawn a great deal of
information from percolation theory ideas \cite{Staufferbook}.
Relatively simple models based on analogies have been implemented
in order to  describe realistic granular pile (static and
dynamic)
properties\cite{reposeangle1,reposeangle2,jams,arches1,arches2,arches3}.
Segregation \cite{segregation}, decompaction
\cite{decompactation} and avalanches \cite{avalanches}
point out to the existence of clusters. However, in   describing such
materials it is crucial to consider that they are not made of
symmetrical  entities. It is necessary to introduce at least one degree
of freedom for the grain with some coupling to an external field.  One
can imagine that the degree of freedom, is called a ''spin'', coupled to 
a ''magnetic field'',  though the ''spin'' can represent any
non-magnetic physical  feature of practical interest, like the grain
roughness or shape feature. The spin role is to break the spatial
isotropic symmetry.  The direction of such a spin can represent the
position of a grain with respect to neighboring entities, as well as a
rotation process. The grain-grain interaction can be imagined to be some
(elastic-like) potential containing information on the grain Young,
rigidity, bulk  modulus, and Poisson ratio, ..   and geometric aspects
\cite{Rosas}. Generalizations to more complex spin models are 
immediately imagined.

Coniglio and Herrmann presented in \cite{conher} a related view of the
granular packing problem and adapted the Ising model and
Sherrington-Kirkpatrick spin glass model to granular phenomena obtaining
two phase transitions in  the system. The short range exchange energy
$J$ describing a ''spin-spin interaction'' is analogous in granular
matter to the contact energy   between grains. An interpretation of $J$
for flows can be found in Pandey et al.\cite{pandey}. A constrained
Ising  spin chain has also been recently considered and studied as a toy
model for granular compaction\cite{toymodel}.

Something which is fully appreciated is the difference in constructing
piles in presence or not of vertical walls. A pyramidal pile has not
necessarily the same structure, arches, ... as a pile in a silo.
Moreover it is quite unrealistic to  build a pile from a single grain
faucet. Finally, due to its anisotropic shape a grain can rotate during
its fall e.g. in responding to the local wind, or difference in pressure
between the upper and lower grain surfaces. Such remarks have motivated
us into re-examining a magnetic Tetris-like model (or a magnetic rain)
for which the grain is characterized by a ''spin'' which can flip during
its fall under some energetic condition.

In this paper the role of changing the depositing spin flipping
probability during its fall in such a MBD (magnetic ballistic depositon)
model is shown to influence the pile density, the pile
''magnetization'', and the cluster size distribution. In
Sect.\ref{sec:experimental}, we establish the algorithm rules and
briefly comment upon them. In Sect.\ref{sec:numerical} we present
numerical results for the density and the magnetization of the pile.
(Sect.\ref{sec:percolation}-\ref{sec:mass}). A critical percolation line
is found separating two regimes for the size (mass) distribution of
clusters. Finally, in Sect.\ref{sec:conclusions}, a brief conclusion can
be found.

\section{ EXPERIMENTAL PROCEDURE} \label{sec:experimental}
The algorithm for the so called $q$-MBD model, in contrast to the
$\frac{1}{2}$-MBD model
\cite{our},   goes as follows:

\begin{enumerate} \item first, we choose a horizontal substrate of spins with a
predetermined (for example antiferromagnetic-like or random) configuration;
periodic boundary conditions are used;

\item{\label{item:step2}} a falling (up or down) spin is dropped 
along one of the
lattice columns from a height $r_{max}+5a$, where $r_{max}$ is the largest
distance between an occupied cluster site and the substrate,

\item at each step down the spin can flip i.e., change its
''$sign$''; the ''up'' direction has a probability $q$;

\item the spin goes down flipping until it reaches a site perimeter of the
cluster at which time  the local gain in the Ising energy

\begin{equation}
\beta E = -\beta J \sum_{<i,j>} \sigma_i \sigma_j
\end{equation}

is calculated. The fall velocity is irrelevant and there is no
backscattering. If the gain is negative the  spin sticks to the cluster
immediately (sticking probability =1.0) and one goes  back to step
(\ref{item:step2}). In the  opposite case the  spin sticks to the
cluster  with a rate $\exp (-\Delta \beta E)$ where $\Delta \beta E$ is
the local gain in the Ising energy. If the spin does not stick to the
cluster it continues going down. Of course if the site just below the
spin is occupied the spin  immediately stops and sticks to the cluster.

\item After the spin stops one goes back to step (\ref{item:step2}).
\end{enumerate}

In the $\frac{1}{2}$-MBD model\cite{our}  a finite field $H$ and 
$q=0.5$ were assumed; in the present report we take $H=0$ but enlarge
the permissible  values of $q$ to  $ [0,1]$.

\section{ NUMERICAL RESULTS} \label{sec:numerical}

All results reported below are for a triangular lattice of
horizontal size $L=100$, when the pile made of clusters has reached a $100$
lattice unit height, and after averaging over $1000$  simulations.
The  substrate consists of spins with random direction.

\subsection{ Density} \label{sec:density}

We define the density of a cluster as $\rho = \frac{\mbox{number
of spins in the cluster}}{\mbox{number of sites on the lattice}}$
in which obviously the number of lattice sites  in the denominator = 10 000.

\begin{figure}
\begin{center}
\includegraphics[height=8cm,angle=-90]{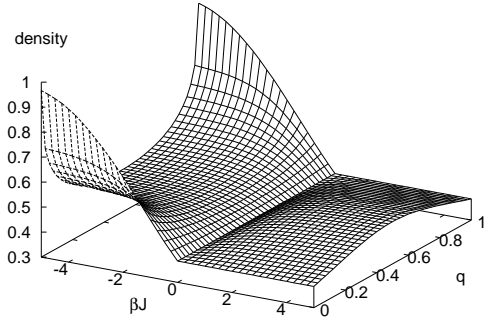}
\end{center}
\caption{\label{fig:density3d} The dependence of the density
on $q$  and $\beta J$}
\end{figure}

Fig.\ref{fig:density3d} illustrates the behavior of the density with
respect to the $q$ and $\beta J$ parameters. This figure convinces us
that the results are symmetrical with respect to $q=0.5$. In the
$\frac{1}{2}$-MBD model the density varies between $0.38$ and $0.47$ but
in the present $q$-MBD a spread in density occurs, -- from almost a
completely compact  pile in the AF case and large $q$ to a loosely
packed pile in the F case. The lowest density (0.38) occurs for  $\beta
J$  =0 and for border  values of $q$, i.e. $0$ and $1$, in the F-case.
For strong enough positive interactions ($\beta J>4$) the density
saturates toward the  value $\rho \approx 0.47$, in the F-case. In the
AF case, the density varies between 0.38 and 1.0.

\subsection{ Magnetization}

\begin{figure}
\begin{center}
\includegraphics[height=8cm,angle=-90]{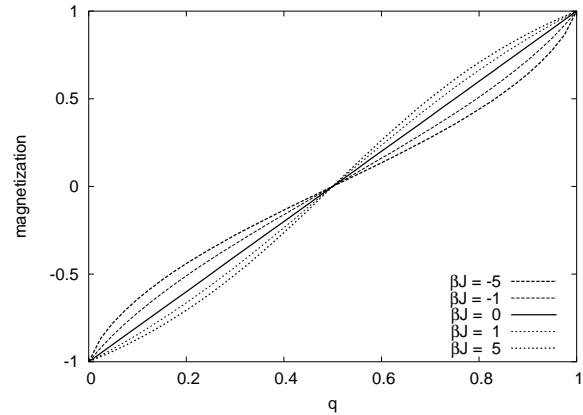}
\end{center}
\caption{\label{fig:magnetization} Dependence of the magnetization on $q$ for
several $\beta J$ values:  F case (dotted lines),
AF case (dashed lines)}
\end{figure}

The dependence of the magnetization defined as

\begin{equation} M = \frac{n_+ - n_-}{n_+ + n_-} \end{equation}

is shown in Fig.\ref{fig:magnetization} as a function of $\beta J$ and $q$,
where $n_+$ and $n_-$ are the number of up and down spins respectively,
i.e. $n_+ = 10000 \rho_+$. $M$ can be considered as a measure
of the difference in grain orientations in  the final packing.

The  magnetization appears to be the same in the regions $(\beta
J>0,q<0.5)$ and $(\beta  J<0,q>0.5)$ (see Fig.\ref{fig:magnetization}). Notice
that there is no  terrace observed here in contrast to the $\frac{1}{2}$-MBD
model case \cite{our}.

\section{ PERCOLATION } \label{sec:percolation}

\begin{figure}
\begin{center}
\includegraphics[height=8cm,angle=-90]{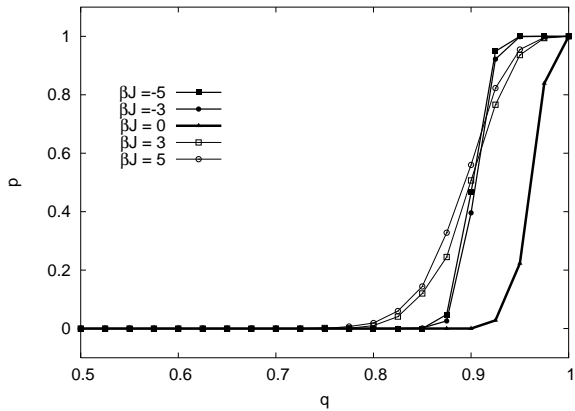}
\end{center}
\caption{\label{fig:percolation2d} Dependence of the percolation probability
$p_c$ on as a function of $q$ for different $\beta J$ values.}
\end{figure}

\begin{figure}
\begin{center}
\includegraphics[height=8cm, angle=-90]{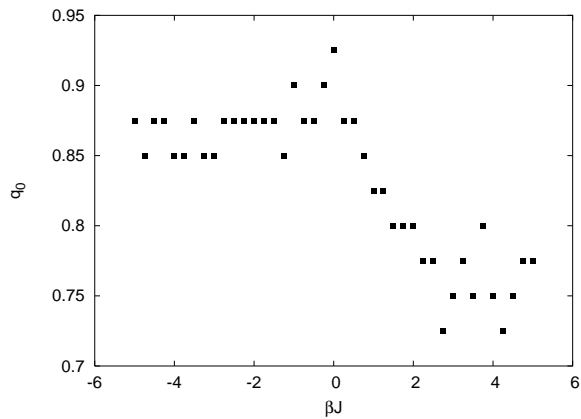}
\end{center}
\caption{\label{fig:p0} Dependence of $q_0$ on $\beta J$.}
\end{figure}

Define an up-spin percolating cluster as a cluster of up spins which
extents from the bottom to the top of the sample. At fixed $q$ and
$\beta J$ the fraction $p$ of piles consisting in such a percolating
cluster of up spins was computed. The behavior of $p$ with respect to
$q$ for several values of the $\beta J$ parameter is shown in
Fig.\ref{fig:percolation2d}. There is no percolating cluster for
$q<0.75$ in the F-case and 0.85 in the AF-case. The $q_0$ ($q_0(\beta J)
:= q $ when $p>0$) dependence is shown in Fig.\ref{fig:p0}. Above $q_0$
when $p$ becomes finite a very fast growth of $p$ is observed, at a
given $q_c$.

The differences between AF and F are  understood  if one recalls that
the density in the AF case is much more sensitive  to a change in  $q$
than in the F case, for which  a  change of $q$ induces only a very
small variation of the density -- recall that the density in this case
is  $[0.38;0.47]$. On the other hand a similar variation in $q$ values
generates piles with a wide interval of densities in the AF case.

\section{ THE SIZE (MASS) DISTRIBUTION OF CLUSTERS} \label{sec:mass}

\begin{figure*}
\begin{center}
\includegraphics[height=16cm, angle=-90]{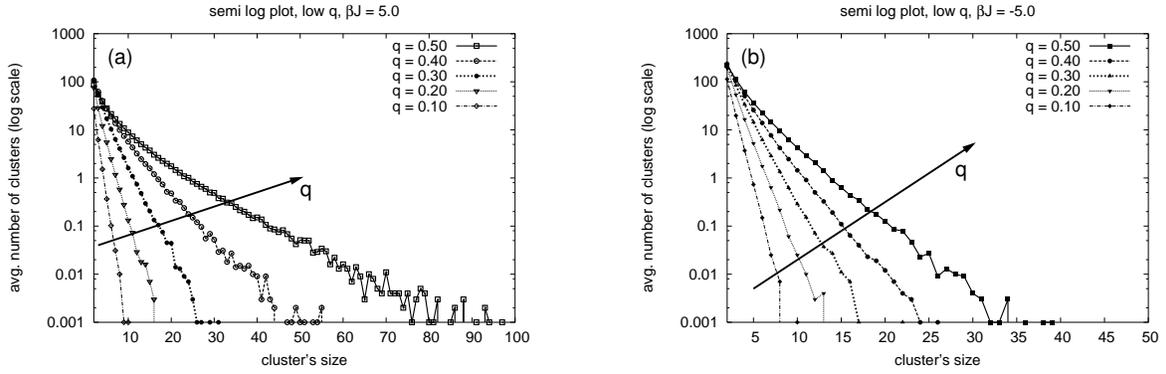}
\end{center}
\caption{\label{fig:low} Examples of the $N_s(s)$ dependence for low
values of  $q$   on semi-log plots: (a) $\beta J$
=5; (b) $\beta J$ =-5}
\end{figure*}

The number of clusters with $s$ size is called $N_s$. Its behavior is
shown in Fig.\ref{fig:low} for low $q$ values. Notice that the $N_s$
dependence is approximately  a straight line  on a semi-log
plot (all logs are neperian), i.e.

\begin{equation}
N_s(s) \propto e^{-k_Es} \qquad \mbox{for} \qquad q \approx 0,
\end{equation}

where $k_E$ is a constant for a fixed $(\beta J,q)$ pair and $s$ is the
cluster size (mass). Observe that the existence of a large
cluster is more probable in the  F case than in the AF case. This is
expected when one recalls that the F case favours neighbors with a
similar  spin direction during the deposition.

\begin{figure*}
\begin{center}
\includegraphics[height=16cm,angle=-90]{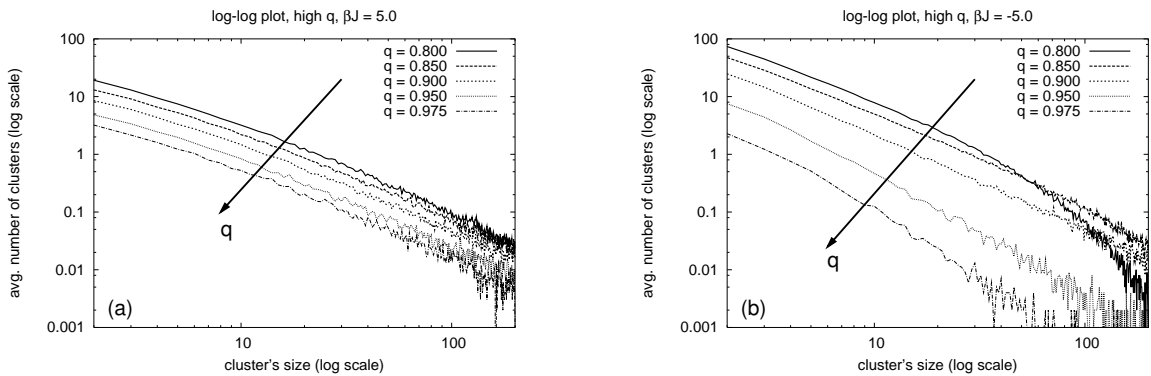}
\end{center}
\caption{\label{fig:high} Examples of the $N_s(s)$ dependences for
high values of the $q$ parameter presented  on log-log plots; (a)
states for ferromagnetic-like case, i.e. $\beta J$ =5; (b) states for
antiferromagnetic-like case, i.e. $\beta J$ =-5 }
\end{figure*}

For high $q$ values the $N_s$ dependence stops to be an exponential law
and becomes a power law (see log-log plots on Fig.\ref{fig:high}). Let
us postulate

\begin{equation}
N_s(s) \propto s^{-k_P} \qquad \mbox{for} \qquad q \approx 1.
\end{equation}

In order to find the values of $q$ for which the crossover effect
occurs, i.e. from an exponential law regime (ELR) to a power law regime
(PLR), we have estimated the slopes of the $N_s$ dependencies in both
regimes. It seems that one should distinguish pile growth conditions
with respect to $q(\beta J)<q_c$ and $q(\beta J)>q_c$.

Analogously to $N_s$ we can define $N_h$ as  the number of hole
clusters.   Unlike the spin clusters there is no difference in $N_h$
dependencies between low and high values of  $q$, -- all $N_h$
dependencies seem to follow a power law.

\section{ CONCLUSIONS} \label{sec:conclusions}

We have presented the extension of the MBD model \cite{our} in order  to
find the role of the probability of  spin flip during deposition (or the
degree  of freedom modification) in granular downward rain-like flow in
2D. The generalized model, hereby called $q$-MBD, is a nonequilibrium
ballistic deposition model with one degree of freedom. One can imagine
that the spins (grains) have  different shapes and $q$ can be related to
a wind strength such that the grains favour one or another position in
order to minimize the pile energy. We have examined the cluster
properties through two order parameters, since two characteristic fields
($J$ and $q$)  are intrinsic to the model.

We have investigated the size, or ''mass'',  of the spin clusters
created through simulation of the nonequilibrium deposition. The
''quenching'' of the degree of freedom on the cluster leads to two
different regions of spin cluster geometric properties. In the low
''field'' $q$ region the spin cluster mass distribution follows an
exponential law, while in the high $q$ region the distribution is
characterized by  a power law. The transition between these two regimes
is not sharp. The exponential law regime for high strength of
interactions (i.e. $|\beta J| > 4$) seems to be universal, i.e.
independent on the magnitude of the intrinsic parameters but depends 
only on the sign of $\beta J$, i.e. the characteristic contact
interaction potential, -- if they are ferromagnetic-like or
antiferromagnetic-like, mechanically  repulsive or attractive.

\begin{acknowledgments}
KT is supported through an Action de Recherches
Concert\'ee Program of the University of Li$\grave e$ge (ARC 02/07-293).
\end{acknowledgments}

\end{document}